\documentclass[aps, prb, twocolumn, notitlepage,superscriptaddress,nofootinbib]{revtex4-1}
 \usepackage{natbib}
\usepackage{graphicx}
\usepackage{dcolumn} 
\usepackage{bm}
\usepackage{hyperref}
\usepackage{mathtools}
\usepackage{amssymb}
\usepackage{enumitem}
\usepackage[mathlines]{lineno}
\usepackage[titletoc, title]{appendix}

\begin{document}

\preprint{APS/123-QED}

\title{Mott-Enhanced Exciton Condensation in a Hubbard bilayer}

\author{Samuele Giuli}
    \email[Correspondence email address: ]{sgiuli@sissa.it}
    \affiliation{International School for Advanced Studies (SISSA), via Bonomea 265, 34136 Trieste, Italy}
\author{Adriano Amaricci}
    \affiliation{CNR-IOM, via Bonomea 265, 34136 Trieste, Italy}
\author{Massimo Capone}
    \affiliation{International School for Advanced Studies (SISSA), via Bonomea 265, 34136 Trieste, Italy}
    \affiliation{CNR-IOM, via Bonomea 265, 34136 Trieste, Italy}

\date{\today} 

\begin{abstract}
We study the conditions to realize an excitonic condensed phase  in an electron-hole bilayer system with local Hubbard-like interactions at half-filling, where we can address the interplay with Mott localization.
Using Dynamical Mean-Field Theory, we find that an excitonic state is stable in a sizeable region of a phase diagram spanned by the intra-layer ($U$) and inter-layer ($V$) interactions. The latter term is expected to favour the excitonic phase which is indeed found in a slice of the phase diagram with $V >U$. Remarkably, we find that when $U$ is large enough, the excitonic region extends also for $U > V$ in contrast with naive expectations.
The extended stability of the excitonic phase can be linked to in-layer Mott localization and inter-layer spin correlations. Using a mapping to a model with attractive inter-layer coupling, we fully characterize the condensate phase in terms of its superconducting counterpart, thereby addressing its coherence and correlation length.

\end{abstract}

\keywords{Exciton, Exciton Condensation, Multi-Orbital Hubbard, DMFT}

\maketitle
\section{Introduction} \label{sec:introduction}

The condensation of excitons in a macroscopic quantum state has been
proposed soon after the success of BCS theory of 
superconductivity\citep{Keldysh,Lozovik} owing to the similarities 
between the Cooper pairs created by the binding of two electrons, and
the excitons, bound states formed by an electron and a hole. However,
the observation of excitonic phases has long eluded the experimental
effort, mainly because of the short lifetimes of the excitons due to
electron-hole recombination processes.

The developments in the engineering of devices and heterostructures
have provided ideal platforms to observe exciton condensation (EC),
which has been indeed proposed and reported in quantum-Hall bilayers
\citep{Spielman,Eisenstain}, graphene double
bilayers\citep{Burg,Li,Perali2013,Amelio2023} and semiconductors
quantum wells \citep{High,Butov}. Excitonic ordering has also
been recently reported also in bulk solids \citep{Cercellier2007,Kogar2017,Afonso2017,Moyoshi2018,Kunes2014,Windgatter2021,Jia2022NP,Sun2022NP}

Bilayer structures are arguably ideal platforms to observe condensation of spatially indirect excitons composed by holes and electrons belonging to different layers, for which recombination is essentially inhibited by the presence of a dielectric material between the layers.
Quantum Monte Carlo calculations for electron-hole gases coupled  by the long-range Coulomb interaction\citep{DePalo1,LopezRios2018,DePalo2023} have indeed shown that an excitonic phase is stable at very low densities, a result which has been confirmed by simulations of double bilayer graphene\citep{Burg,Li}. 


In an analogous lattice model with local interactions  some indication of exciton condensation has been found away from half-filling\cite{Rademaker} and in the half-filled system when the interlayer interaction is larger than the intra-layer repulsion\citep{Huang1,Huang2}.
Similar models have  been investigated using Dynamical Mean-Field
Theory (DMFT).
In Ref.~\onlinecite{Vanhala} the competition between EC and  s-wave superconductivity has been addressed in a model without intra-layer repulsion. 
A variety of two-orbital models including, e.g., energy splitting between bands, the  Hund's coupling and including non-trivial topology have also been found to host excitonic states in some regions of parameters\citep{Kaneko,Kunes0,Kunes1,Kunes2,Niyazi2020,Amaricci2023}.

In this work we aim at identifying a generic mechanism 
connecting strong correlation physics and excitonic phases which can be used to gain a deeper insight on results on more involved and richer models for specific systems. In particular,
we address the interplay between the EC and Mott physics, the most direct fingerprint of correlations, in an idealized model for an electron-hole bilayer system with local Hubbard-like interactions.
Our focus is on the relative role of the intra-layer ($U$) and
inter-layer ($V$) interactions. We consider the system at
half-filling, where a Mott transition can take place, so that our phase diagram will be characterized by the competition/interplay between Mott insulating and EC phases.

The paper is organized as follows: In Sec. II we introduce the model, our implementation of Dynamical Mean-Field Theory and the relevant observables we consider. In Sec. III we present the normal-phase results where we discard excitonic ordering, while Sec. IV is devoted to the results for the EC phase. Sec. V reports our concluding remarks.

\section{Model and Method} \label{sec:modelandmethod}
We consider a two-layer Hubbard model with a local interaction term: 
\begin{eqnarray} 
    H  & =- \sum_{\langle ij \rangle \sigma m} t_m c^\dagger_{i \sigma
         m} c_{j \sigma m} + H.c. -\mu \sum_{i \sigma m} n_{i \sigma m}\nonumber
    \\
&+U \sum_{i m}n'_{i \uparrow m}n'_{i \downarrow m}  +V \sum_{i \sigma \sigma^\prime}n'_{i \sigma A} n'_{i \sigma^\prime B} 
\label{eq:ham}
\end{eqnarray}
where $c_{i \sigma m}$ ($c^\dagger_{i \sigma m}$) is the annihilation (creation)  operator of an electron on site $i$, layer $m=A,B$ and with spin $\sigma$, $n_{i \sigma m}$ is the number operator and $n'_{i \sigma m} = n_{i \sigma m}-1/2$ is introduced to write the model in a particle-hole symmetric form which implies that both bands are half-filled for $\mu =0$. 
We set $t_A = t$ and $t_B=\alpha t_A$. In our calculations we will consider $\alpha = -1$ in order to describe an electron-like band (A) and a hole-like band (B). 
$U$ and $V$ are both positive and they measure the intra-layer and inter-layer local screened Coulomb repulsion.

We will study an excitonic state characterized by a uniform ($q=0$) spin-singlet excitonic order parameter (EOP)
\begin{equation}
    \Delta_0 = \frac{1}{N} \sum_{i\sigma}  \langle c^\dagger_{i A \sigma}c_{i B \sigma} \rangle \ 
\end{equation}
which is expected to be degenerate with spin-triplet counterparts due to the SU(2)$\times$SU(2) spin symmetry of our model. Models including other interaction terms and material-specific features, may favour one or the other spin symmetries\citep{Kunes1,Kunes2,Amaricci2023}.



We solve the model at zero temperature using DMFT\cite{DMFT_GKKR}, a state-of-the-art method which treats different interactions non perturbatively and it is particularly well suited to study the Mott transition\cite{DMFT_GKKR}, strongly correlated metallic phases as well as superconductivity and other broken-symmetry states. Within DMFT the lattice model is mapped onto an impurity model which has to be solved self-consistently requiring that the impurity Green's function coincides with the local component of the lattice Green's function. We solve the impurity model at $T=0$ using 
Lanczos/Arnoldi  exact diagonalization (ED)\cite{Caffarel,CaponeED,amaricci2021edipack}. 
As customary in the DMFT community, we consider a Bethe lattice with a semicircular density of states  $N_m(\epsilon)= \frac{2}{\pi D_m^2} \sqrt{D_m^2-\epsilon^2 }$, where $D_m \propto t_m$ is the half-bandwidth.

In order to study the EC phase, the bath of the impurity model has to include an excitonic amplitude, analogously to the superconducting case.
Using a spinorial representation where $\Psi_{k,\sigma}^\dagger =( c_{k \sigma A}^\dagger , c_{k \sigma B}^\dagger )$, where $k=0$ identify the impurity and $k= 1,...,N_{bath}$ the bath levels, we can write it as 

\begin{align} 
H_{imp}^{(0)} =& \sum_{k \sigma} \begin{pmatrix}\Psi_{k \sigma}^\dagger & \Psi_{0 \sigma}^\dagger \end{pmatrix}
\begin{pmatrix}
\mathcal{H}_{k \sigma} & V_k \cdot \mathbb{I}_{2} \\ V_k \cdot \mathbb{I}_{2} & 0
\end{pmatrix}
\begin{pmatrix} \Psi_{k \sigma} \\ \Psi_{0 \sigma} \end{pmatrix}\label{H_AIM_full_su2}
\end{align}

where $\mathbb{I}_2$ is the  $2 \times 2$ identity and
\begin{align}
\mathcal{H}_{k \sigma} =& \begin{pmatrix}
\epsilon_{k} + M_k & P_k \\
P_k & \epsilon_{k} -M_k \\
\end{pmatrix}  
\label{bath_localhamiltonian}
\end{align}

where $P_k$ is the inter-orbital excitonic hybridization term in the
bath Hamiltonian, $\epsilon_k + (-) M_k$ is the bath energy on orbital
$A$ ($B$) and $V_k$ is the hybridization between the impurity and bath
site $k$.
Within ED-DMFT we have to limit the number of bath sites to be able to solve the impurity model.  We fixed the number of bath sites to be $N_{bath}=4$ and we fixed the system at global half-filling $\langle \sum_{ \sigma m } n_{\sigma m} \rangle = 2$ by imposing $\mu=0$, then since we are focusing on orbitals with opposite dispersion relation we also fixed $\epsilon_k=0 \ \ \forall k$ and since we focus on state with orbital half-filling, this required that for each $M_k$ parameter on bath site $k$ there must be another bath site $k^\prime$ with opposite energy $M_{k^\prime}=-M_k$. 


\section{Normal State} \label{subsec:mottinsulatingstates}
We start our investigation from the normal state where we inhibit excitonic ordering, as well as any other broken-symmetry state  like antiferromagnetism or staggered orbital ordering.  This is a standard strategy which has helped to understand the Mott transition disentangling Mott localization from magnetic ordering\citep{DMFT_GKKR}. For our model, a normal-state phase diagram has been reported in Ref. \citep{Koga3}, but we find it useful to present our results in order to emphasize the aspects which are useful to better address the excitonic phase. 

The model is expected to feature two different Mott-insulating solutions that we can easily understand from the atomic ($t_m =0$) limit. Among all configurations with two electrons per site, the four with one electron in each layer $\vert\uparrow,\downarrow\rangle$, $\vert\downarrow,\uparrow\rangle$, $\vert\uparrow,\uparrow\rangle$ and $\vert\downarrow,\downarrow\rangle$ have energy $E_{11} = -\frac{1}{2} U$, while the two configurations with two electrons in the same layer $\vert\uparrow\downarrow,0\rangle$ and 
$\vert 0,\uparrow\downarrow\rangle$ have energy $E_{20} = \frac{1}{2} U-V$. Therefore the former set of states is favoured for $U > V$ and the latter for $U < V$. 
Hence when $U$ and $V$ are much larger than the hopping and $U > V$ we expect an insulator with one electron on every site of each layer.
This state, that we label as U-Mott (U-MI) is expected to be unstable towards antiferromagnetic ordering if we allow for symmetry breaking. On the other hand, for $V>U$ we have an insulator where every site is in a mixture between the two solutions with one doubly occupied layer. This state, henceforth V-Mott (V-MI),  would be naturally unstable towards a staggered orbital (layer) ordering.

    \begin{figure}[h!]
        \centering
        \includegraphics[width=0.45\textwidth]{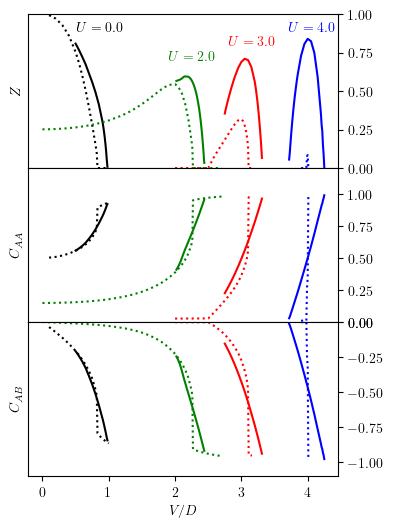}
        \caption{Quasiparticle weight (top), intra-orbital density-density correlation  (center) and inter-orbital density-density correlation (bottom), as a function of $V/D$ for $U/D=0.0$ (black), $2.0$ (green), $3.0$ (red) and $4.0$ (blue). Dotted lines are data in the normal state, solid lines mark the same quantities in the excitonic phase}
        \label{fig:ZCC}
    \end{figure}
In order to monitor the Mott localization we compute the quasiparticle
weight $Z_m$ which measures the metallicity of the
system\citep{DMFT_GKKR}.
The progressive destruction of the metallic state is  described by a
reduction of $Z_m$ from 1 (non-interacting limit) to 0 (correlated
insulator).
The  connected local density-density correlations $C_{m,m^\prime} =
\langle n_m n_{m^\prime} \rangle - \langle n_{m}\rangle\langle
n_{m^\prime}\rangle$ can be used to study the competition between the
two {\bf interaction} terms and the approach to the atomic
limit insulators.
The orbital symmetry implies $C_{AA} = C_{BB}$ and $C_{AB} = C_{BA}$.
It is easy to see from the above discussion that 
the atomic $U-MI$ has $C_{AA}=0$ and $C_{AB}=0$, while the atomic
$V-MI$ has $C_{AA}=1$ and $C_{AB}=-1$. 

    \begin{figure}[h!]
        \centering
        \includegraphics[width=0.4\textwidth]{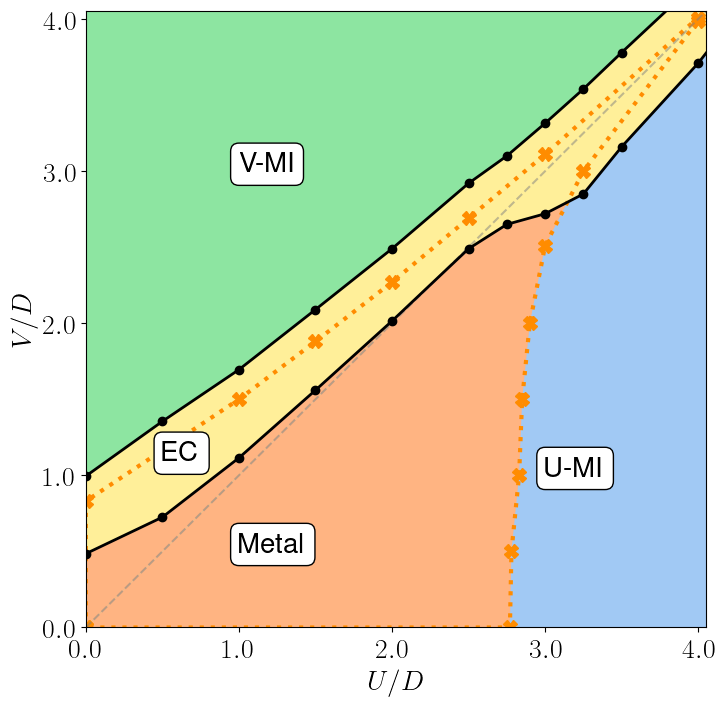}
        \caption{
          $V$ vs $U$ Ground State Phase Diagram. In \textit{yellow}
          the region of EC phase, in \textit{orange} the metallic
          phase, in \textit{blue} the U-Mott insulator and in
          \textit{green} the V-Mott one.
          The dashed lines with crosses symbols indicate  the two
          Mott-transition boundaries in the normal state, while the gray dashed line highlight the $U=V$ line.}
        \label{fig:Full_PhaseDiag}
    \end{figure}
    
In Fig. \ref{fig:ZCC} we show as dotted lines the evolution of
$Z_A=Z_B$ and of the inter- and intra-layer correlations $C_{AA}$ and
$C_{AB}$ as functions of $V/D$ for different values of $U/D$.
 The boundaries of the U-MI and V-MI phases are marked by dotted lines
 with crosses  in the phase diagram of Fig. \ref{fig:Full_PhaseDiag}

 The cuts for $U/D =$ 1 and 2 in Fig. \ref{fig:ZCC}  clearly show a
 metal-insulator transition towards  the V-MI state with $Z_A=0$,
 $C_{AA}=1$ and $C_{AB}=-1$. For $U/D =$ 3, we find a U-MI for small
 $V$ followed by a metallic region and the V-MI as $V$ increases.
 For large $U/D =4$ we have only a tiny slice of $V$ with a metallic solution 
 sandwiched by the two insulators.

The main feature of the normal-state phase diagram, as already pointed
out in Ref. \onlinecite{Koga3}, is the existence of a metallic region when $U$
and $V$ are comparable, even when they are so large to independently drive a Mott
transition (in the absence one of the other). The region shrinks as we
increase $U$ and $V$ but it does not close.
In particular, for $U=V$ we always find a metallic solution, similarly
to other models where the competition between different atomic states
leads to intermediate phases which can have either a
metallic\cite{Isidori,Richaud} or an insulating\cite{Scazzola}
nature.

\section{Excitonic Phase} \label{subsec:excitoncondensation}


    \begin{figure}[h!]
        \centering
    \includegraphics[width=0.5\textwidth]{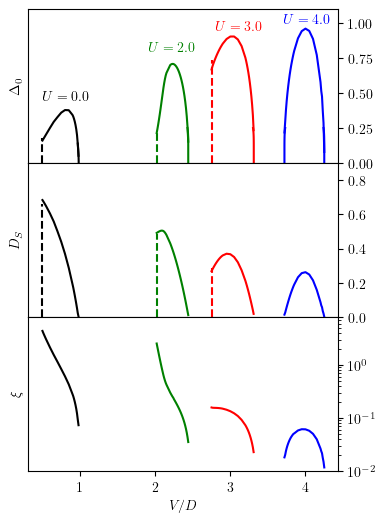}
    \caption{Excitonic order parameter $\Delta_0$ (top), stiffness $D_s$(center) and coherence length $\xi$ for (from left to right) $U/D=0.0$, $2.0$, $3.0$, $4.0$ with the same color codes of Fig. 1. The vertical dashed line indicate the first order Metal-EC phase transition.}
    \label{fig:dat_EOP_Ds}
    \end{figure}

We now turn to solutions where the exciton condensation is allowed. The values of $Z_A$, $C_{AA}$ and $C_{AB}$ are shown as solid lines in Fig. \ref{fig:ZCC} and compared with their normal-state counterparts. Indeed, the excitonic state is stable in a wide region of parameters and its onset makes the evolution from the U-MI to the V-MI smoother, thereby increasing also the quasipartcle weight.

 Reporting this information on the phase diagram of
 Fig. \ref{fig:Full_PhaseDiag}, where the boundaries of the excitonic
 region are black solid lines,  we clearly see that the EC region is
 roughly centered around the normal state transition towards the
 V-Mott state. The picture is simple: Increasing $V$, before the
 interaction is large enough to drive the system insulating, it leads
 to the binding {\bf of electrons and holes on different layers into} excitons.
 However, the effect of $U$ changes the
 position and the nature of the transition.
 
For small and moderate $U$ the EC establishes only when $V$ prevails over $U$ (above the $V=U$ line, marked with a dashed grey line) in agreement with previous work\citep{Huang1,Huang2,Vanhala}.

A much less expected result emerges when we increase $U$ and we approach the boundary of the U-MI phase. Here we find that the stability region of the EC increases and, remarkably, it extends in the region where $U < V$ signaling a non-trivial intrinsic many-body effect due to the interplay of the two interactions. As a result, for $U \gtrsim 3D$, the whole metallic region between the two Mott insulators is replaced by an excitonic state.

The positive effect of the Hubbard repulsion on the excitonic order is evident in Fig. \ref{fig:dat_EOP_Ds} (a), where we plot the order parameter $\Delta$ as a function of $V$ for the same cuts of Fig. 1.
Here we show that the EC for large $U$ is not only stable in a wider range of $V$, but its amplitude is also larger. For instance, for $U/D=4$ the maximum value of $\Delta$ is more than twice the $U=0$ maximum. For every value of $U$, the transition from the metal to the EC appears of first-order, while the transition from the EC to the V-MI state is associated with a continuously vanishing $\Delta$.

\subsection{Exciton Ordering and Mott physics} \label{subsec:magneticcorrelations}



In this section we link the enhancement of the EC region  for $V<U$ and large $U/D$ to the magnetic correlation between orbitals near the V-MI phase that is enhanced by the nearby U-MI phase.
The main effect of $U$ is to drive a standard Mott localization within
each layer. Hence the double occupation on each layer $d_m$ is
strongly reduced.
For a half-filled non-magnetic system this reflects directly in the formation of local moments as measured by $\langle S_m^z S_m^z \rangle = \frac{1}{4}\langle (n_{m,\uparrow} -n_{m,\downarrow})^2 \rangle = \frac{1}{2}(\frac{1}{2} -d_m)$ which approaches 1/4.
While the spins on the two layers are uncorrelated in the normal state, when we reach the EC region and $U \gtrsim 3D$ the inter-layer spin correlations  $\langle S^z_A S^z_B \rangle$ become sizeable and negative eventually approaching the limit -1/4. 

    \begin{figure}[h!]
        \centering
        \includegraphics[width=0.5\textwidth]{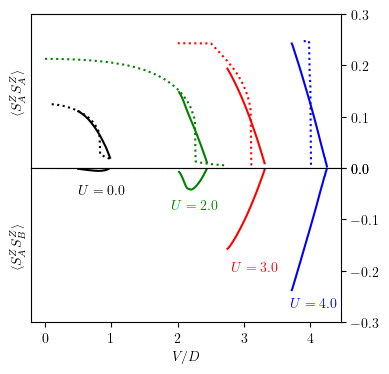}
        \caption{ Local magnetic moments (intra-orbital spin
          correlations) (top) and inter-orbital magnetic correlation
          (bottom). Dotted and solid lines indicate, respectively, the
          normal and the excitonic phase solution.
          Data are for $U/D=0.0$ (black), $2.0$ (green), $3.0$ (red) and $4.0$ (blue).}
        \label{fig:dat_spinspin}
    \end{figure}
The local quantum state (computed from the impurity model within DMFT) approaches for large $U$
$| \psi \rangle \sim \frac{1}{\sqrt{2}}(|\uparrow_A \downarrow_B
\rangle + |\uparrow_B \downarrow_A \rangle )$ for which $\langle S^z_A
S^z_A \rangle =\frac{1}{4}$ and $\langle S^z_A S^z_B \rangle
=-\frac{1}{4}$.

Note however that the interplay between Mott localization and exciton ordering is not trivial. The singlet atomic excitonic state is indeed a linear combination of  
$|\uparrow_A \downarrow_B \rangle$ and  $|\uparrow_B \downarrow_A \rangle$ which are favoured by increasing $U$, but also of the states $|\uparrow_A \downarrow_A, 0 \rangle$ and
$|0, \uparrow_B \downarrow_B \rangle$, which are instead depleted by $U$. Hence, while the magnetic correlations develop approaching the U-Mott state,
they first contribute to the onset of excitonic ordering, but as we exceed a given "optimal" distance from the Mott state, the EOP decreases, leading to the existence of a bell-shaped behavior of the order parameter.

We finally notice that  the spin-singlet correlations follow from our choice to study spin-singlet excitons, and we expect the same picture to hold for spin-triplet exciton. The key idea is that Mott localization within each layer leads to localized moments which are naturally prone to acquire any inter-layer correlation when exciton ordering is allowed. Finally, in the U-MI state the EOP vanishes and the $SU(2)\times SU(2)$ spin symmetry with four independent ground states is recovered.

\subsection{Characterizing the Excitonic State via a mapping on Superconductivity} \label{subsec:superfluidity}

A particle-hole transformation on layer B:
\begin{equation}
\label{eq:particle-hole}
c^\dagger_{i \sigma B} \rightarrow c_{i \sigma B} (-1)^{\sigma}
\end{equation}
maps our model for $\alpha $= -1 onto  a two-orbital model with the same form of Eq. (1) in which  the two orbitals share the same hopping $t_A = t_B = t$ and the inter-orbital interaction becomes attractive (-$V$), while the intra-layer remains repulsive. This model can indeed host an inter-orbital s-wave superconducting state, which maps on our excitonic state via the same particle-hole transformation (\ref{eq:particle-hole}).
We can exploit this mapping to compute some observable which characterize the superconducting state and allow to better characterize the EC. 

The superfluid stiffness $D_s$ \citep{Scalapino} is a crucial parameter that controls the critical temperature. It measures the coherence of the superconducting state and its rigidity to fluctuations of the phase of the order parameter. Indeed, a  superconductor with small $D_s$ has a small critical temperature even if the zero-temperature modulus of the order parameter is large, as it happens in the strong-coupling limit in a single-orbital attractive Hubbard model \citep{Toschi}
In the effective model with inter-layer attraction $-\vert V\vert$ obtained via the transformation (\ref{eq:particle-hole}) $D_s$ reads
\begin{equation}
    \frac{D_S}{\pi e^2}= \langle -E_{kin} \rangle - \chi_{jj} ( \mathbf{q}\rightarrow 0 , \omega =0)
\end{equation}
where $j$ is the current operator and $E_{kin}$ is the expectation value of the hopping part of the Hamltonian.
For a Bethe lattice we obtain\citep{Toschi}
\begin{equation}
    \frac{D^{ex}_S}{e^2 \pi} = -\frac{4 \alpha}{\beta} \sum_{ i \omega_n , \sigma}  \int d\varepsilon  V(\varepsilon) D(\varepsilon) |G_{AB} (\varepsilon , i\omega_n)|^2
\end{equation}
where $V(\epsilon)=\frac{4t^2-\epsilon^2}{2}$ is the square of the current vertex for orbital $A$ and $\alpha=t_B/t_A$  (See Appendix \ref{sec:appendix_superfluidstiffness} for derivation) . 

 We underline that the total current of the attractive model corresponds, in model (\ref{eq:ham}), to the operator
\begin{equation}
j_{ex}(\mathbf{q},i\omega_n)=j_A(\mathbf{q},i \omega_n)- j_B(\mathbf{q},i \omega_n),
\end{equation}
which is clearly different from the current operator associated with the total charge. Hence, the $D_s$ can be considered a real superfluid stiffness only for the auxiliary attractive model.

Yet, $D_s$ provides direct also information about the coherence and stability properties, which translates into an analogous information about the EC phase of our model (\ref{eq:ham}).

The coherence length $\xi$ has indeed naturally the same meaning in the two frameworks, namely it measures the length over which the constituents of the pair/exciton retain quantum coherence. It is given by\citep{Seki,Kaneko2}
\begin{equation}
    \xi^2 = \frac{ \sum_\mathbf{k} | \nabla_{\mathbf{k}} F( \mathbf{k}) |^2}{\sum_\mathbf{k} | F( \mathbf{k}) |^2} 
\end{equation}
where
\begin{equation}
    F(\mathbf{k}) = \sum_{i\omega_n} e^{i\omega_n 0^+} G_{AB}(\epsilon_{\mathbf{k}},i\omega_n)
\end{equation}

The results for $D_s$ and $\xi$ are reported in panels (b) and (c) of
Fig. \ref{fig:dat_EOP_Ds} in order to compare their behavior with the
EOP. The results for $U=0$ are qualitatively similar to an attractive
model and they reflect the BCS to Bose-Einstein Condensate (BEC) crossover as a function of the coupling. Indeed both $D_s$ and $\xi$ are maximal in the weak-coupling side and they decrease as the interaction grows. 

Increasing $\vert V\vert$ we have a progressive reduction of the coherence length, associated with more localized pairs/excitons characteristic of the BEC limit. Also $D_s$ decreases as result of the smaller coherence of the pairs/excitons and it actually vanishes at the continuous transtion to the V-MI state.

When we introduce and increase $U$, we find an important difference on the "weak-coupling" side of the crossover. Indeed both $D_s$ and $\xi$ are depleted also close to the smallest values of $V$ required to establish the EC. As a result, for large $U$ the two quantities have a maximum around the $U \sim V$ line. These results clearly confirm the $U$-induced localization of the excitons that we discussed above and the crucial role of the interplay between the two interactions to induce an EC for $V<U$.

\section{Conclusions} \label{sec:conclusions}

We used DMFT to assess the existence of  an excitonic state in the zero-temperature phase diagram of a two-layer Hubbard model with intra-layer ($U$) and inter-layer ($V$) density-density repulsive interactions. Working at half filling, we can study how the excitonic long-range order is affected by the Mott physics. 

We find a sizeable region of exciton ordering when the two interactions are comparable.
The transition from EC phase to the Mott insulating phase is continuous, while the transition from Metal to EC is of the first order.

For small and intermediate $U$, the excitonic state is present only if $V > U$. On the other hand, for $U \gtrsim 3D$ i.e., close to a standard Mott transition within each layer, we find an exciton state also when $V < U$, signaling a non-trivial interplay in which quantum fluctuations play an active role.

We have indeed shown that the enlargement of the excitonic phase in the proximity of the intra-layer Mott transition can be connected with the $U$-driven development of local magnetic moments that, in turn, favour magnetic correlations between the two layers (singlets in our case). We expect this mechanism to be general, and in particular, to be present also for models where the exciton and the magnetic correlations  have a triplet symmetry. 

Exploiting a simple mapping onto a model with attractive inter-layer
interactions, we have been able to further characterize the excitonic
state. The coherence length, which has essentially the same
interpretation of that of a superconductor, shows that the proximity
to the V-driven Mott state leads to localized pairs with very short
coherence length. Analogously, the equivalent of the superconducting
superfluid stiffness shows that the coherence of the EC state tends to
vanish when the V-Mott insulator is reached. In other words, when we
approach the Mott transition, the EC state is driven towards the
strong-coupling limit, which in the superconducting language corresponds to the BEC limit\citep{Toschi}. We notice in passing that the
BEC nature and its evolution from a BCS limit can be experimentally
assessed via both thermodynamic\citep{Toschi} and spectral
properties\citep{Sangiovanni2006AFM,Taranto2012}.
These results further strengthen our picture where the charge
localization induced by $U$ is central in the stabilization of the
excitonic condensate for $V < U$ and in determining its properties.

The existence of excitonic states for $V < U$ is important because in a real bilayer system, or in a multi-orbital correlated material, we always expect $V < U$. We notice however that an electron-phonon coupling of the Holstein type (coupled with the total local electron  density) can effectively reduce $U$, making in principle the effective $U$ closer or even smaller than $V$\citep{Sangiovanni2005,Sangiovanni2006,Scazzola}.

As we anticipated in the introduction, our model has been introduced as the minimal model for a bilayer system in which excitonic phases can be present and, at the same time, Mott physics is effective. The results we have obtained have to be considered as a basis to build the understanding of richer and more involved models including, among others, different and more complex hopping structures, energy difference and/or hybridization betweeen the two bands and a richer structure of the interactions.

\section*{Acknowledgements} \label{sec:acknowledgements}
We acknowledge funding by MUR through the PRIN 2017 (Prot. 20172H2SC4
005), PRIN 2020 (Prot. 2020JLZ52N 002) programs, National Recovery and
Resilience Plan (NRRP) MUR Project No. PE0000023-NQSTI and ICSC–Centro
Nazionale di Ricerca in High Performance Computing, Big Data and
Quantum Computing, funded by European Union – NextGenerationEU (Grant
number CN00000013) -  Mission 4 Component 2 Investments 1.3 and 1.4.

\bibliography{biblio}

\appendix

\section{Superfluid Stiffness} \label{sec:appendix_superfluidstiffness}

In this appendix we provide some details of the calculation of the superfluid stiffness for the attractive model obtained through the canonical transformation (\ref{eq:particle-hole}). From the definition\citep{Scalapino}:
\begin{equation}
    \frac{D_S}{\pi e^2}= \langle -E_{kin} \rangle - \chi_{jj} ( \mathbf{q}\rightarrow 0 , \omega =0)
\end{equation}
We need to compute the kinetic energy and the current-current response function. We make use of the previously defined spinorial representation to define the Green's function as:
\begin{widetext}
\begin{equation} \label{app:spinorial_G}
\hat{G}_\sigma (\mathbf{k},\tau)= \langle T \begin{pmatrix}
c_{k A \sigma}(\tau) \\ c_{k B \sigma}(\tau)
\end{pmatrix} \otimes
\begin{pmatrix}
c^\dagger_{k A \sigma}(0) & c^\dagger_{k B \sigma} (0) 
\end{pmatrix} \rangle =
\begin{pmatrix}
G_{AA} (\mathbf{k} , \tau ) & G_{AB} (\mathbf{k} , \tau )  \\
G_{BA} (\mathbf{k} , \tau ) & G_{BB} (\mathbf{k} , \tau )  
\end{pmatrix}
\end{equation}
\end{widetext}
From now on we consider it diagonal in the spin therefore we can avoid to write explicitly the spin index $\sigma$. In single-site DMFT, where the self-energy is local and site independent, the Dyson equation for the interacting Green's functions reads:
\begin{equation} \label{eq:app_dyson}
    \hat{G}_0 (\mathbf{k},i \omega_n )^{-1} = \hat{G} (\mathbf{k},i \omega_n )^{-1} + \hat{\Sigma} ( i \omega_n )
\end{equation}
where the hat indicates that all of these are matrices as in the previous equation \ref{app:spinorial_G}. This means that the diagonal and off diagonal component are:
\begin{widetext}
\begin{align}
G_{AA} ( \varepsilon , i \omega ) =& \frac{ i \omega - \alpha \varepsilon -\Sigma_{BB} ( i \omega)}{\big( i\omega - \varepsilon - \Sigma_{AA}(i \omega) \big) \big( i \omega -\alpha \varepsilon -\Sigma_{BB}(i \omega) \big) - | \Sigma_{AB}(i \omega) |^2  }  \\
G_{BB} ( \varepsilon , i \omega ) =& \frac{ i \omega - \varepsilon -\Sigma_{AA} ( i \omega)}{\big( i\omega - \varepsilon - \Sigma_{AA}(i \omega) \big) \big( i \omega -\alpha \varepsilon -\Sigma_{BB}(i \omega) \big) - | \Sigma_{AB}(i \omega) |^2  }  \\
G_{AB} (\varepsilon, i \omega ) =& \frac{ \Sigma_{AB} ( i \omega)}{\big( i\omega - \varepsilon - \Sigma_{AA}(i \omega) \big) \big( i \omega -\alpha \varepsilon -\Sigma_{BB}(i \omega) \big) - | \Sigma_{AB}(i \omega) |^2  }=G_{BA}^* (\varepsilon, i \omega )
\end{align}
\end{widetext}
where $\alpha=t_{B}/t_{A}$  therefore $\epsilon^{(A)}=\epsilon$ and $\epsilon^{(B)}=\alpha \epsilon$. 

In this derivation we will set the energy splitting to zero ($M=0$) for simplicity but the results remain valid for any value of $M$. In DMFT the kinetic energy for orbital $m$ can be easily computed since the Green's function is known:
\begin{align}
E_{kin}^{(m)} =& \sum_{\mathbf{k} \sigma} \epsilon_{\mathbf{k}}^{(m)} \langle c^\dagger_{\mathbf{k} \sigma m} c_{\mathbf{k} \sigma m} \rangle \nonumber \\
\end{align}
\begin{align}
=& \lim_{\eta \rightarrow 0^{+}} \beta^{-1} \sum_{i \omega_n}\sum_{\mathbf{k} \sigma} \epsilon_{\mathbf{k}}^{(m)} G_{m m} (\mathbf{k},i \omega_n ) e^{i \omega_n \eta} \nonumber \\
=& \lim_{\eta \rightarrow 0^{+}} \beta^{-1} \sum_{i \omega_n, \sigma} \int d \epsilon D(\epsilon) \epsilon^{(m)} G_{m m} (\epsilon,i \omega_n ) e^{i \omega_n \eta}
\end{align}
computing it explicitly for the two orbitals and performing a partial integration using the relation $-\epsilon D(\epsilon)=\partial_\epsilon [ D(\epsilon) V(\epsilon)]$ where $V(\epsilon)=\frac{4t^2-\epsilon^2}{3}=(v^{(A)}_{\epsilon})^2$ is the square of the current vertex in orbital $A$, $\alpha^2 V(\epsilon)=(v^{(B)}_{\epsilon})^2$ is the square of the current vertex in orbital $B$ and $D(\epsilon)= \frac{1}{2\pi t^2}\sqrt{(2t)^2-\epsilon^2}$ is the density of states:
\begin{widetext}
\begin{align} \label{ekin}
E_{kin,A} =& \beta^{-1}\sum_{ i \omega_n , \sigma}  \int d\varepsilon \ V(\varepsilon) D(\varepsilon) G_{AA}^2 (\varepsilon , i\omega_n) \Big[1 +\alpha \frac{|\Sigma_{AB}(i \omega_n) |^2}{\big(i \omega_n - \alpha \varepsilon - \Sigma_{BB}(i \omega_n) \big)^2 } \Big]  \nonumber \\
=&\beta^{-1}\sum_{ i \omega_n , \sigma}  \int d\varepsilon \ V(\varepsilon) D(\varepsilon) \Big[ G_{AA}^2 (\varepsilon , i\omega_n)+\alpha |G_{AB} (\varepsilon , i\omega_n)|^2\Big] \\
E_{kin,B} =& \beta^{-1}\sum_{ i \omega_n , \sigma}  \int d\varepsilon \ V(\varepsilon) D(\varepsilon) G_{BB}^2 (\varepsilon , i\omega_n) \Big[\alpha^2 +\alpha \frac{|\Sigma_{AB}(i \omega_n) |^2}{\big(i \omega_n -  \varepsilon - \Sigma_{AA}(i \omega_n) \big)^2 } \Big]  \nonumber \\ 
=& \beta^{-1}\sum_{ i \omega_n , \sigma}  \int d\varepsilon \ V(\varepsilon) D(\varepsilon)  \Big[\alpha^2 G_{BB}^2 (\varepsilon , i\omega_n)+\alpha |G_{AB} (\varepsilon , i\omega_n)|^2 \Big] \\
\end{align}
\end{widetext}
From which one can check that if there is no orbital off-diagonal self-energy and $\alpha=\pm 1$ the kinetic energy is the same in the two orbitals. The computation of the current-current response in DMFT in infinite dimensions is simplified since all the vertex corrections are cancelled \citep{DMFT_GKKR} and only the elementary bubble contributions survive, therefore:
\begin{widetext}
\begin{align}
    \chi_{jj} (\mathbf{q} , \tau) \ =& \ -\langle j_{ex} (\mathbf{q}, \tau) j_{ex}(-\mathbf{q},0) \rangle \ , \ j_{ex}(\mathbf{q}, \tau)=j_{A}(\mathbf{q}, \tau)-j_{B}(\mathbf{q}, \tau) \\
    \chi_{jj} (\mathbf{q} \rightarrow 0 , i \omega =0) \ =& \ [\chi^{AA}_{jj} - \chi^{AB}_{jj} - \chi^{BA}_{jj} +\chi^{BB}_{jj} ](\mathbf{q} \rightarrow 0 , i \omega =0) \\
    \chi_{jj}^{mm^{\prime}} ( \mathbf{q},i\omega) \ =& \ -\beta^{-1} \sum_{\mathbf{k} , i\nu , \sigma} v^{(m)}_{\mathbf{k} \sigma} v^{(m^{\prime})}_{\mathbf{k}+\mathbf{q} \sigma} G_{mm^{\prime}}( \mathbf{k}, i\nu) G_{m^{\prime} m} (\mathbf{k} + \mathbf{q}, i \nu + i \omega) \ \ , \ \ m,m^{\prime}=A,B  \\
\end{align}
\end{widetext}
Where the current vertex for the two orbitals are related by $v^{(B)}= \alpha v^{(A)}$.
Merging the DMFT results for the kinetic energy and the current-current response function, the superfluid stiffness for the selected model is:
\begin{widetext}
\begin{equation}
    \frac{D_S}{e^2 \pi} = -\frac{4\alpha}{\beta}\sum_{ i \omega_n , \sigma}  \int d\varepsilon \ V(\varepsilon) D(\varepsilon) |G_{AB} (\varepsilon , i\omega_n)|^2 
\end{equation}
\end{widetext}
This interesting result carries some important information. Since the \textit{Superfluid Stiffness} has to be a positive quantity, the "naive" two-orbital Hubbard model with symmetric bands ($\alpha=1$) would not allow any finite $D_S$, this is in agreement with some results showing that local excitonic correlations are dumped for $\alpha>0$  \citep{Pavol2} in favor of a bipartite antiferro-EC state that correspond to a model with a shift of the $B$ band of the vector $\mathbf{Q}$ of bipartite lattices for which $\epsilon_{\mathbf{k}} = - \epsilon_{\mathbf{k}+\mathbf{Q}}$, e.g. for the square lattice in $D$-dimensions the vector is $\mathbf{Q}=(\pi,\pi,...,\pi)$. For $\alpha=0$ (Falikov-Kimball Model with spin) it correctly predict no superfluid excitonic state since one of the species is not mobile and since in this limit no excitonic phase is expected \citep{Pavol}. This special case prohibit excitonic ordering since in the limit $\alpha \rightarrow 0^+$ there must be an antiferro-EC state while in the limit $\alpha \rightarrow 0^-$ a ferro-EC state, thus $\alpha=0$ is an unstable point between these two
phases\citep{Kunes2}.
Our choice of opposite bands $\alpha=-1$ is therefore optimal and in this situation the Superfluid Stiffness can be rewritten as:
\begin{equation}
    \frac{D_S}{e^2 \pi} = \frac{4}{\beta} \sum_{\sigma , i\omega_n} \int d  \varepsilon  \, V ( \varepsilon ) D( \varepsilon ) | G_{AB}( \varepsilon , i\omega_n) |^2
\end{equation}
This results tells us that opposite band dispersion is the optimal ground for the research of a Superfluid Exciton Condensate.

\section{Calculation of the Coherence Length} \label{sec:appendix_coherencelength}
For the Bethe lattice we have no access to the momenta but only energy, therefore we have to pass from $\nabla_\mathbf{k}$ to something we can treat. Starting from the numerator of the coherence length definition\cite{Seki,Kaneko2}:
\begin{widetext}
\begin{equation}
    \sum_{\mathbf{k}} \Big|\nabla_{\mathbf{k}} F( \mathbf{k})\Big|^2 
    =\sum_{\mathbf{k}} \Big|(\nabla_{\mathbf{k}} \epsilon_{\mathbf{k}})\frac{\partial F(\epsilon)}{\partial \epsilon} \big|_{\epsilon=\epsilon_\mathbf{k}} \Big|^2
    = \sum_{\mathbf{k}} \Big|(\nabla_{\mathbf{k}} \epsilon_{\mathbf{k}}) \big[ \frac{1}{\beta}\sum_{i\omega_n} e^{i\omega_n 0^+}\frac{\partial }{\partial \epsilon}F(\epsilon , i\omega_n) \big|_{\epsilon=\epsilon_\mathbf{k}}\big] \Big|^2,
\end{equation}
\end{widetext}
where $F(\epsilon,i\omega_n)=G_{AB}(\epsilon,i\omega_n)$ as previously defined (See Appendix \ref{sec:appendix_superfluidstiffness}) and $\nabla_{\mathbf{k}}\epsilon_\mathbf{k}=v_{\mathbf{k}}$ is the group velocity of the non interacting particles (we take $\hbar =1$). Now the dependency on $\mathbf{k}$ is present only through $\epsilon_\mathbf{k}$ via the relation $|v_\mathbf{k}|=\sqrt{\frac{4t^2-\epsilon_{\mathbf{k}}^2}{3}}=v(\epsilon)$ therefore we can pass to the integral in energy and the result for the numerator is:
\begin{widetext}
\begin{equation}
    \sum_{\mathbf{k}} \Big|\nabla_{\mathbf{k}} F( \mathbf{k})\Big|^2 
    = \int d\epsilon \ D(\epsilon) \Big| \frac{1}{\beta} \sum_{i\omega_n} e^{i\omega_n 0^+}v(\epsilon) G_{AB}^2(\epsilon , i\omega_n)\frac{2\epsilon +\Sigma_{BB}(i\omega_n)-\Sigma_{AA}(i\omega_n)}{\Sigma_{AB}(i\omega_n) } \Big|^2
\end{equation}
\end{widetext}
For the denominator no change is needed and the substitution of $F(\mathbf{k})$ gives directly
\begin{widetext}
\begin{equation}
\int d\epsilon \ D(\epsilon) \Big| \frac{1}{\beta} \sum_{i\omega_n}e^{i\omega_n 0^+} G_{AB}(\epsilon, i\omega_n)   \Big|^2    
\end{equation}
\end{widetext}

\end{document}